\documentclass[aps,twocolumn,groupedaddress,showpacs,pra]{revtex4-1}

\usepackage{amsmath}    
\usepackage{amsfonts}
\usepackage{bm}
\usepackage{amssymb}
\usepackage{graphicx}   

\begin{document}

\title{Steady state ab-initio theory of lasers with injected signals}


\author{Alexander Cerjan}
\author{A.~Douglas Stone}
\email[]{douglas.stone@yale.edu}
\affiliation{Department of Applied Physics, Yale University, New Haven, CT
06520, USA}

\date{\today}

\begin{abstract}
We present an ab-initio treatment of the steady-state of lasers with injected signals that
describes a regime, valid for micro lasers, in which the locking transition is dominated
by cross-saturation and spatial hole-burning. The theory goes beyond standard approaches
and treats multimode lasing with injected signals and finds the possibility of partially locked 
states and as well as repulsion of the free-running frequencies from the injected signal.
The theory agrees well with exact integration of the full wave and matter equations for the system.
It can also describe accurately complex modern lasers structures and is applied to the example of
deformed disk lasers. We show that in the case of a one dimensional cavity in the locked or
regenerative amplifier regime the theory reduces to an improved version of the Adler equations in the appropriate limit.
\end{abstract}


\maketitle

\section{Introduction}

Laser action in the presence of an injected signal is an extremely important topic for both research in non-linear
dynamics and laser physics, and for applications of lasers. Under certain conditions, a laser mode can be
locked to the injection frequency, allowing for stabilization and modulation of a ``slave" laser based on control by a ``master" laser. 
This and related effects have been the topic of a large literature going back to the beginning of laser theory
\cite{pantell65,tang67,boikova67,lamb72} and certain features are now well-described in 
textbook treatments \cite{siegman}, where the basic paradigm is that of frequency locking of non-linear oscillators as described by Adler \cite{adler46} well before the invention of the laser.  

From the mid seventies onward \cite{haken}, it was also appreciated that lasers with injected signals can exhibit complex dynamical behavior and even chaos based on the general principle
of non-linear dynamics that damped driven non-linear systems with three or more independent time-varying fields 
generically have nontrivial dynamics over large regions of phase space \cite{ohtsubo,wieczorek_review_05}. Since the
basic laser equations involve three distinct and possibly complex functions (the electric field, the polarization and the inversion) 
a self-oscillating laser without injection can exhibit this behavior \cite{haken}, but in most lasers the time scales are such 
that the polarization field (Class B) or the polarization and inversion (Class A) can be adiabatically eliminated leaving only one or two independent field(s).  

Class B lasers are by far the most common type (they include semiconductor and most other solid state lasers), 
and so injecting an additional signal can in many cases generate interesting dynamical states outside the locking region. 
Hence Class B lasers with an injected signal have been used extensively to study such states in the past thirty years.
Deterministic chaos was first reported in Class B lasers with injected signals by Arecchi \textit{et al.}\ \cite{arecchi_ABC_84}, who
introduced the Class A,B,C categories based on the relative size of the 
decay rates of the electric field, $\kappa$, polarization, $\gamma_\perp$, and
inversion of the gain medium, $\gamma_\parallel$. (Class C is the case in which all time constants are comparable, 
none can be adiabatically eliminated, and the laser in isolation can exhibit chaotic dynamics. 
These lasers are very rare and are not technologically important.) The goal of much of this earlier work has been to understand and 
categorize all of the different regimes of stability, bistability, and instability for injected class B lasers \cite{arecchi_ABC_84,mogensen85,oppo_pushing_86,solari_94,zimmermann_global_01,kelleher_BtoA_12}.
These interesting dynamical effects arise because the injected signal beats against the existing, free-running laser 
output, and drives the inversion to oscillate at this beat frequency. When this frequency coincides with other 
relevant dynamical scales in the laser, usually the frequency of the relaxation oscillations, $\omega_r \sim 
\sqrt{\kappa \gamma_\parallel}$, resonant driving occurs leading to complex dynamics.
In addition, for semiconductor lasers, there are dynamical scales associated with the dispersion of the gain medium \cite{lang82},
and to carrier dynamics, which enter the equations as well; these effects arise from the 
real part of the gain susceptibility at the lasing frequency, and are quantified by the Henry $\alpha$ factor \cite{lang82,henry82,henry86,mogensen85}. 

In the current work we will develop a theory of injection locking of Class A and B lasers in a regime relevant particularly to 
micro lasers, in which complex dynamical states do not arise, and for atomic-like gain media for which the $\alpha$ factor is negligible.  
The existence of such a regime does not seem to have been clearly identified in previous work on injection 
locking.  In this regime the physical effect is primarily that of quenching of the free-running laser oscillation due to cross-saturation.  
There have been some relatively recent works \cite{even97,lariontsev99} emphasizing locking through cross-saturation as opposed to synchronization, 
but these models do not include spatial hole-burning, which we find to play an important role, 
and the models also involve many
more approximations than our method, which is essentially an exact solution of the problem in the relevant regime.  

Our approach is a generalization of Steady-state Ab initio Laser Theory (SALT) which has previously been shown to provide 
extremely accurate solutions of the multimode laser equations, again in a regime relevant particularly to micro lasers.
The accuracy of the generalized theory, Injection-SALT, or I-SALT, is here confirmed by direct simulation of the relevant Maxwell-Bloch 
equations describing the laser with injected signal; we have not found any previous works which have performed such simulations either to check a model or as stand-alone results.
In the limit where the laser is locked to the input signal we show that an approximate treatment of our theory reduces to an
Adler type of steady-state solution, but that the behavior outside of the locking range is completely different than expected from
the Adler description. Moreover the Adler approximation is not very accurate for the phase difference between
the locked input signal and the resulting amplified output.

To present I-SALT it is useful to review briefly SALT and the assumptions underlying that theory.
SALT is based on a multiperiodic ansatz for the steady-state solutions of the semiclassical 
Maxwell-Bloch equations describing lasing.  After transient effects have subsided it is assumed that
the electric field in the laser cavity is a sum of some number, $N_L$, of unknown lasing modes, $\Psi_\mu (x)$,
with unknown lasing frequencies, $\omega_\mu$, which are found by solving self-consistent coupled 
non-linear frequency-domain wave equations (see below). These equations include space dependent saturation of the gain
and mode competition to infinite order.  
The number of lasing modes is not assumed known, but is also determined self-consistently 
from the theory, which predicts the thresholds including the effects of mode competition. The theory treats
the openness of the cavity exactly through the introduction of a non-hermitian basis set of outgoing functions, 
which are termed the threshold constant flux basis (TCF), and which will play a crucial role in I-SALT.  
The theory is also formulated for arbitrary cavity geometries and mode functions, so it is well suited to treat
novel modern microlasers such as those based on disk, toroid, photonic crystal or random cavities \cite{mccall_1992,armani_toroid_2003,chua11,jiang04}.
SALT is easily generalized to N-level lasing with a single lasing transition (not mode), and with minor complications
can be applied to multiple lasing transitions and more complex gain media, as long as the basic approximation 
underlying the theory holds. The only substantial approximation in SALT is the neglect of the field beating terms 
in the multimode regime which can lead to complex dynamics and destabilize the multimode solution. 
This approximation leads to an inversion density which varies in space but not in time, hence we
refer to it as the stationary inversion approximation (SIA).
By its nature,  SALT and its generalization to I-SALT, 
will not describe complex dynamical effects in injected lasers of the type mentioned above. 
 
However, as noted, there is an interesting regime in which such effects do not occur, and in which
SALT and I-SALT will describe accurately the steady-state lasing or lasing with injection 
in either the locked or unlocked state. The theories will predict
fully the classical fields, their frequencies, output power, emission pattern etc., except properties due 
to quantum fluctuations, such as the linewidth.  However, recent extensions of SALT \cite{chong12,pick} have
found exact linewidth formulas which are generalizations of Schawlow-Townes based on SALT solutions.
SALT without injection has been compared to full FDTD simulations of microlasers using 
the Maxwell-Bloch and N-level semiclassical lasing
equations, and has been found to agree very well, with a much reduced computational overhead.  For higher
dimensional structures and the full vector Maxwell equations SALT can be used where FDTD is computationally
impractical \cite{cerjan12,esterhazy14}.

\section{Validity of SALT and I-SALT}

We now address the validity of the SIA which will define the regime of validity of multimode SALT and of
I-SALT.  A number of works in early laser theory rely on the SIA, including a seminal paper by Spencer and
Lamb \cite{lamb72} which derives the Adler equations for the injected laser from the Maxwell equations with
injection into a cavity with uniform gain.  Most relevant to our work is that of Fu and Haken in 1991 \cite{haken91}, 
who argued that the SIA was valid and 
steady-state multimode operation was possible as long as $\gamma_\parallel, \kappa \ll \gamma_\perp, \Delta$, where $\gamma_\perp$ 
is the relaxation rate of the polarization and $\Delta$ is the free spectral range of the laser.  They then studied a 
simplified model of a Fabry-Perot type laser and showed that the multimode state with the largest number of modes
was typically the stable state.  They also pointed out that in order for $\Delta \gg \gamma_\parallel$ to hold one typically
would need to look at linear laser cavities of length $L \leq 100 \mu \rm{m}$.  

Fu and Haken did not justify the requirement 
$\kappa \ll \gamma_\perp$ in their work and we find that through comparison with FDTD simulations the SIA and SALT work well even when $\kappa \geq \gamma_\perp$
(``bad cavity" limit).
In the latter case the polarization cannot be adiabatically eliminated and does not follow the electric field instantaneously 
until steady-state is reached; but in the steady-state the SIA holds and the lasing fields are accurately determined by
SALT.  $\kappa$ itself is not a relevant frequency scale for the validity of the SIA in the laser without injection; as long 
as $\gamma_\parallel \ll \gamma_\perp, \Delta$ the SIA, and hence SALT, will describe the steady-state.  
As noted above, the relaxation oscillation frequency, $\omega_r \sim \sqrt{\kappa \gamma_\parallel}$, can be relevant if
it coincides with beat frequency of nearby modes, i.e.\ it is $\sim \Delta$, so that relaxing fluctuations could be resonantly
enhanced and destabilize the multimode state.  However, since $\kappa \leq \Delta, \omega_r \leq \sqrt{\Delta 
\gamma_\parallel} < \Delta, \gamma_\perp$; (we assume that in the interesting cases $\gamma_\perp \geq \Delta$, 
otherwise multimode lasing is unlikely, since $\gamma_\perp$ is the width of the gain curve).  So for steady-state 
multimode lasing without injection all that is required for our free-running theory to work is 
$\gamma_\parallel \ll \gamma_\perp,\Delta$.  

For the injected laser the inversion beat frequency is not $\Delta$, but the
frequency difference between the injected signal $\omega_{in}$ and the free-running signal $\omega_1$ (we assume here
only one free running mode and one injected signal for simplicity). For the generalization to I-SALT to work in the unlocked
regime, where there are two beating signals, we must have $\omega_{in} - \omega_1 > \omega_r$.  
However, even if this is not the case, I-SALT will still describe quantitively the locked regime and predict the unlocking threshold exactly.

In the cases studied below, in which I-SALT describes both the locked and unlocked behavior, we 
find a novel effect in 1D Fabry-Perot type cavities: 
instead of the free-running frequency being ``pulled in" to the injected frequency, as in
the standard Adler picture \cite{siegman}, instead we find that the lasing frequency is repelled from the
injected signal frequency due to the effects of gain competition and spatial hole-burning.  To our
knowledge this behavior is not predicted in any previous works. In Oppo \textit{et al.}\ frequency
repulsion is also found in a certain limit, but it is due to the dynamical effects of relaxation oscillations which are absent for the
cases we consider and thus is a distinct effect.  Moreover, essentially all of the injection literature treats single-mode
one-dimensional cavities. I-SALT naturally allows the description of multimode lasing with
injection, leading to the possibility of a partially locked lasing state, in which one or more modes have been quenched
by cross-saturation, while other modes still lase, as we will demonstrate below.  Also, I-SALT provides a formulation
for describing injection into a cavity with arbitrary two or three-dimensional geometry; we will apply the method to
injection into a 2D chaotic cavity laser below.

The outline of this paper is as follows, in section III we will derive I-SALT using a convenient basis in which
to express the problem, which also facilitates its solution numerically. In section IV we derive a version of the 
Adler steady-state theory from I-SALT, valid for relatively high Q cavities.  In section V we will present
numerical results comparing I-SALT to direct integration of the Maxwell-Bloch equations in time and demonstrate
excellent agreement in the regime of relatively low-Q cavities where we expect other approaches to fail. 
This section will also demonstrate the frequency repulsion of the lasing mode from
the injected mode, in contrast to predictions of previous theories.  We will also present here a comparison between
the I-SALT version of the Adler equation in the locked regime and full I-SALT for higher Q cavities. And finally, we will
present an application of I-SALT to injection locking in two-dimensional cavities. In Section VI we present a restricted stability
analysis of the I-SALT solutions. In section VII we will summarize and make some concluding remarks.

\section{Derivation of I-SALT}

As introduced above, SALT was formulated to determine directly the steady-state of the laser rate equations for an N-level
atomic gain medium coupled to Maxwell's equations within an arbitrary cavity specified by its passive dielectric function, a
tensor in general,
$\varepsilon_c(\mathbf{x})$, and subject to a spatially-varying pump, $D_0 (\mathbf{x}) = d_0 F(\mathbf{x})$, without performing time integration to steady-state. 
Assuming stationary level populations (stationary inversion for the two-level medium), a multimode steady state
exists and is described by a set of time-independent wave equations coupled through their non-linear saturation terms, and 
subject to the non-hermitian boundary condition of purely outgoing solutions at the lasing frequencies.
The SALT equations are solved efficiently by introducing a specific self-orthogonal basis set of 
threshold constant flux (TCF) states in which to expand the solutions, and then iteratively solving the resulting non-linear 
matrix equation for both fields and frequencies.  We now show how this approach can be generalized to yield I-SALT.

The Maxwell-Bloch equations,
which describe the coupling of the electric field to a population of two level gain atoms \cite{haken},
\begin{align}
  4 \pi \partial_t^2 \mathbf{P}^+ =& c^2 \left( \nabla \times \nabla \times \mathbf{E}^+ \right) - \varepsilon_c(\mathbf{x}) \partial_t^2 \mathbf{E}^+ \\
  \partial_t \mathbf{P}^+ =& -\left(\gamma_\perp + i\omega_a \right)\mathbf{P}^+ + \frac{\mathbf{g}}{i \hbar} D \left(\mathbf{E}^+ \cdot \mathbf{g}\right) \\
  \partial_t D =& -\gamma_\parallel \left(D - d_0 F(\mathbf{x}) \right)  \notag \\
  &- \frac{2}{i \hbar} \left(\mathbf{E}^+ \cdot (\mathbf{P}^+)^* - (\mathbf{E}^+)^* \cdot \mathbf{P}^+ \right),
\end{align}
form the basis for SALT.
In these equations $\mathbf{E}^+$ and $\mathbf{P}^+$ are the positive frequency components of the electric field and atomic polarization
respectively, $\omega_a$ is the atomic transition frequency, and $\mathbf{g} = e \langle e | \hat{\mathbf{x}} | g \rangle$ is the dipole matrix element
in which $| e \rangle$ and $| g \rangle$ are the excited and ground spatial states of the electron wave functions. In separating
the electric field into its positive and negative frequency components, terms oscillating at twice the atomic frequency have been omitted, 
corresponding to the usual rotating wave approximation (RWA). Note that we have not introduced the slowly-varying
envelope approximation for the electric field as is usually done; this is unnecessary for the steady-state and provides no
computational advantage.  As noted above, it has been previously demonstrated that any
N-level atomic medium with a single lasing transition in the steady-state can be reduced to an effective two level
atomic system of the type considered here in the steady-state limit \cite{cerjan12}. 
Furthermore, SALT has also been shown to describe more complex gain media (C-SALT),
which have multiple lasing transitions and diffusion of the gain atoms \cite{CSALT}, as long as the SIA
still holds. However for simplicity in this
manuscript we will focus only on the Maxwell-Bloch equations as written above.

As discussed above, we assume the existence of a steady-state with stationary level populations, which in general requires that
$\gamma_\parallel , \omega_r \ll \delta \omega, \Delta, \gamma_\perp$, where $\delta \omega$ is the detuning
of the injected signal from the free-running laser frequency and $\omega_r,\Delta,\gamma_\perp$ are as previously defined. 
However for a cavity with only a single operating mode, either injected or free-running, this inequality is not necessary, 
and the I-SALT solution is exact (in the RWA).
In general, the positive frequency components of the electric field and atomic polarization inside the 
cavity for a given pump value, $d_0$, take the form
\begin{align}
  \mathbf{E}^+(\mathbf{x},t) =& \sum_\mu^{N_L} \mathbf{\Psi}_\mu(\mathbf{x}) e^{-i\omega_\mu t} + \sum_\alpha^{N_A} \mathbf{\Psi}_\alpha(\mathbf{x})e^{-i \omega_\alpha t} \label{ansatz} \\
  \mathbf{P}^+(\mathbf{x},t) =& \sum_\mu^{N_L} \mathbf{p}_\mu(\mathbf{x}) e^{-i\omega_\mu t} + \sum_\alpha^{N_A} \mathbf{p}_\alpha(\mathbf{x})e^{-i \omega_\alpha t}, \label{ansatz2}
\end{align}
where the $N_L$ lasing modes, $\mathbf{\Psi}_\mu (\mathbf{x})$, and associated polarization fields, $\mathbf{p}_\mu(\mathbf{x})$, have unknown spatial variation, and 
unknown frequencies, $\omega_\mu$, and there are $N_A$ amplified signals injected into the
cavity, at {\it given} frequencies, $\omega_\alpha$, and {\it given} incoming amplitudes, $B_\alpha$, but with unknown 
overall amplitude, spatial variation, $\mathbf{\Psi}_\alpha (\mathbf{x})$, and polarization, $\mathbf{p}_\alpha(\mathbf{x})$, within the cavity.  
All of the unknown quantities will be determined from the resulting I-SALT equations and their boundary conditions self-consistently.  
We now insert the multi-periodic ansatz (\ref{ansatz}-\ref{ansatz2}) 
into the Maxwell-Bloch equations,
and apply the SIA to write
\begin{equation}
\mathbf{p}_\sigma(\mathbf{x}) = \frac{\mathbf{g}}{\hbar} \frac{D(\mathbf{x})}{\omega_\sigma - \omega_a + i \gamma_\perp}  \left(\mathbf{\Psi}_\sigma(\mathbf{x}) \cdot \mathbf{g} \right),
\end{equation}
where $\sigma$ is either a free-running or injected mode.
This allows for the elimination of the polarization and atomic inversion, leading to
$N_L + N_A$ coupled non-linear wave equations, which can be written
as three-dimensional vectorial equations, but which here we will only consider in their scalar form, appropriate for the 
geometries studied below:
\begin{align}
  \left[ \nabla^2 + \left(\varepsilon_c(x) + \frac{\gamma_\perp D(x)}
    {\omega_\sigma - \omega_a + i \gamma_\perp} \right) k_\sigma^2 \right] \Psi_\sigma(x) 
  = 0 \label{salt_field} \\
   D(x) = \frac{d_0 F(x)}{1 + \sum_{\mu}^{N_L} \Gamma_\mu |\Psi_\mu(x)|^2
  + \sum_{\alpha}^{N_A} \Gamma_\alpha |\Psi_\alpha(x)|^2}, \label{ampinv}
\end{align}
where $\Gamma_\sigma \equiv \gamma_\perp^2 / [(\omega_\sigma - \omega_a)^2 + \gamma_\perp^2]$ is the gain curve and 
$k_\sigma = \omega_\sigma/c$
is the wavevector. The electric field
and inversion have also been scaled to natural units, $E_c = 2 g / \hbar \sqrt{\gamma_\parallel \gamma_\perp}$ and
$D_c = 4 \pi g^2 / \gamma_\perp \hbar^2$.
The wave equations for lasing modes, $\Psi_\mu (x)$, are to be solved with purely outgoing boundary conditions, while those
for amplified modes, $\Psi_\alpha (x)$, are to be solved with the boundary condition of fixed input amplitude $B_\alpha$ at $\omega_\alpha$.

We will solve these coupled equations by non-linear iteration after expanding the solutions 
in a non-Hermitian basis set with the appropriate boundary conditions.  For the lasing modes, $\Psi_\mu$, this set is the
same TCF states used in SALT, which satisfy,
\begin{align}
  \left[ \nabla^2 + \left(\varepsilon_c(x) + \eta_n F(x) \right) k^2 \right] 
  u_n(x; \omega) =0 \label{outtcf} \\
  \partial_x u_n(x; \omega)|_{x=L} = i k u_n(L; \omega) \\
  u_n (0;\omega) = 0,
\end{align}
where we refer to $u_n(x; \omega)$ and $\eta_n (\omega)$ as the TCF eigenvectors and eigenvalues respectively and
we have written the outgoing boundary condition explicitly for a one-sided cavity with a perfect mirror at the origin of the type
we study below.  The general outgoing boundary conditions is expressed differently for different geometries, but is well known.
Note that $\{\eta_n (\omega)\}$ is generically complex and can be thought of as the set of values of the gain medium susceptibility
which lead to lasing at frequency $\omega$, i.e.\ for which a solution for purely outgoing real wavevector exists \cite{ge10}.
In this basis the lasing mode is written as
\begin{equation}
  \Psi_\mu(x) = \sum_n a_n^{(\mu)} u_n(x; \omega_\mu). \label{lasedecomp}
\end{equation}
Thus the lasing thresholds in the absence of input signals are obtained by varying $\omega$ in the TCF equation until
a frequency, $\omega_\mu$, is found
at which 
\begin{equation}
\eta_n (\omega_\mu) = \frac{d_0\gamma_\perp}{\omega_\mu - \omega_a + i \gamma_\perp}.
\end{equation}
The $\{ u_n (x;\omega_\mu)\}$ then form an efficient basis set for finding the {\it non-linear} solutions above threshold
because at the first lasing threshold, the lasing mode is only a single TCF state, and above threshold only a 
small number of TCFs are needed to converge to the non-linear solution of the SALT equations.

The amplified modes, $\Psi_\alpha$, however must be treated differently from the lasing modes since they have a fixed incoming signal amplitude 
and fixed frequency. To represent these modes we require terms with an incoming component in addition to
the outgoing TCF expansion terms, which we do conveniently by solving the same TCF equation inside the cavity
with a purely incoming boundary condition,
\begin{equation}
  \Psi_\alpha(x) = \sum_n a_n^{(\alpha)} u_n(x; \omega_\alpha) + 
  \sum_m b_m^{(\alpha)} v_m(x; \omega_\alpha), \label{ampdecomp}
\end{equation}
where the states $v_m(x; \omega)$ and associated eigenvalues $\beta_m$ are  given by
\begin{align}
  \left[ \nabla^2 + \left(\varepsilon_c(x) + \beta_m F(x) \right)k^2 \right]
  v_m(x; \omega) = 0 \label{intcf} \\
  \partial_x v_m(x; \omega)|_{x=L} = -i k v_m(L; \omega) \\
  v_m (0;\omega) = 0,
\end{align}
and thus represent states that are purely incoming. 
The incoming and outgoing TCF states are not power orthogonal, but they do satisfy a self-orthogonality condition between
themselves,
\begin{align}
  \frac{1}{L}\int_C dx F(x) u_n(x; \omega) u_m(x; \omega) = & \delta_{nm} \\
  \frac{1}{L}\int_C dx F(x) v_n(x; \omega) v_m(x; \omega) = & \delta_{nm},
\end{align}
which can be derived from the definitions of the states and Green's theorem.
Either the incoming or outgoing
TCF states represent a complete basis for fields {\it within the cavity} at $\omega_\alpha$, but the incoming
terms are needed to represent the input boundary condition.  Because 
they are purely incoming, they do not contribute directly to the emitted fields, but they correctly represent the full
spatial hole-burning and gain competition effects of the amplified input. 

For amplified modes, we can easily write
the incoming boundary condition for a one-sided slab cavity of length $L$ as
\begin{equation}
  B_{\alpha} e^{-i k_\alpha L} = \sum_m b_m^{(\alpha)} v_m(L; \omega_\alpha), \label{inputdef}
\end{equation}
where $B_\alpha$ is the given incoming field amplitude at frequency $\omega_\alpha$.
This single equation vastly under-determines the coefficients $b_m^{(\alpha)}$ in the sum, so 
that the choice is based on convenience. This freedom arises from the overcompleteness of
using both $\{u_n\}$ and $\{v_m\}$ to represent the internal fields.
Hence the coefficients $a_n^{(\alpha)}$ depend strongly on the choice of the $b_m^{(\alpha)}$.
A natural choice is to take only a single term, $v_0(x; \omega_\alpha)$, which corresponds to the dominant component
of the outgoing TCF state for the nearest lasing mode.  This is allowed for a cavity with a single input
channel, as in the one-sided slab geometry we are considering here; in general one needs a minimum
of $M$ independent incoming states to represent an arbitrary input for an $M$-channel cavity, and these can
be chosen again to be similar in character to the nearest lasing mode in order to optimize the calculation.

Once a representation of the input field is chosen, one can insert the expansions (\ref{ampdecomp})
for the amplified modes and (\ref{lasedecomp}) for the lasing modes into
the fundamental Eqs.\ (\ref{salt_field}) and (\ref{ampinv}) and use the self-orthogonality relations of the outgoing TCF states
to find coupled non-linear matrix equations for the coefficients $a_n^{(\mu)}, a_n^{(\alpha)}$
which determine their solutions.  For the lasing modes, $\Psi_\mu$, one finds
\begin{align}
  \eta_l a_l^{(\mu)} &= \sum_n T_{ln}^{(\mu)} a_n^{(\mu)} \label{IS1} \\
  T_{ln}^{(\mu)} &= \frac{\gamma_\mu d_0}{L}
  \int_C dx \frac{F(x) u_l(x; \omega_\mu) u_n(x; \omega_\mu)}{1 + 
    \sum_\sigma^{(N_L+N_A)} \Gamma_\sigma |\Psi_\sigma(x)|^2}, \label{I-SALT1}
\end{align}
where $\gamma_\mu = \gamma_\perp/(\omega_\mu - \omega_a + i \gamma_\perp)$.  This is identical to 
the lasing equations of SALT except for the presence of the
amplified mode intensities in the non-linear hole-burning denominator. 

In a similar manner the coupled equations for the amplified modes can be determined, and
they take the form:
\begin{align}
  \eta_l a_l^{(\alpha)} &= \sum_n T_{ln}^{(\alpha)} a_n^{(\alpha)} 
  + \sum_m \left(W_{lm}^{(\alpha)} + V_{lm}^{(\alpha)} \right) b_m^{(\alpha)} \label{I-SALT2a} \\
  W_{lm}^{(\alpha)} &= \frac{\gamma_\alpha d_0}{L} \int_C dx \frac{F(x) u_l(x; \omega_\alpha) v_m(x; \omega_\alpha) }
  {1 + \sum_\sigma^{(N_L+N_A)} \Gamma_\sigma |\Psi_\sigma(x)|^2} \label{I-SALT2b} \\
  V_{lm}^{(\alpha)} &= \frac{\beta_m}{L} \int_C dx F(x) u_l(x; \omega_\alpha) v_m(x; \omega_\alpha). \notag
\end{align}
The result for the overlap integral can be simplified further through the use of
the definitions of the incoming and outgoing TCF states and Green's theorem,
allowing one to write
\begin{equation}
  V_{lm}^{(\alpha)} = \frac{2i}{L k_\alpha} \frac{\beta_m}{\beta_m - \eta_l}
  u_l(L; \omega_\alpha) v_m(L; \omega_\alpha) \label{Vdef}.
\end{equation}
As with any basis expansion method, this method of representing the original differential equations will
require truncation of the sums at a finite number of TCFs, $N$, in the numerical implementation.

From this form of the overlap integral, it is simple to understand why the most numerically
efficient choice of incoming TCF states to use, $b_m^{(\alpha)} \ne 0$, are those related to
the outgoing states of the nearest lasing mode. In the case of a lossless cavity, $\varepsilon_c(x) \in \mathbb{R}$,
the incoming and outgoing TCF states form a biorthogonal set, with $\beta_m = \eta_m^*$ and $v_m = u_m^*$. Thus,
$V_{lm}$ is maximized when the difference between the incoming and outgoing TCF eigenvalues
is minimized. This choice allows the outgoing states required by $T_{ln}$ to also have significant
overlap with the incoming states chosen, rather than needing to include additional outgoing states
to properly compute the sum over overlap integrals.

The I-SALT equations for free-running modes (\ref{IS1}-\ref{I-SALT1}) have a critical difference
from the equations for amplified modes (\ref{I-SALT2a}-\ref{Vdef}): in the former,
there is an undetermined global phase whereas for the latter the phase is set by the injected
signal, $B_\alpha$. For lasing modes the undetermined global 
phase is chosen by convention (gauge condition) \cite{tureci07,ge10}.
This leaves $2N-1$ expansions coeficients to fully determine the real and imaginary parts of $a_n^{(\mu)}$,
and one additional equation which determines the unknown lasing frequency. It is this equation
which determines the full intensity dependent line-pulling effects on the lasing frequencies, and,
in the case of I-SALT, frequency-pulling or pushing due to the injected mode. In contrast, for the amplified mode the
frequency and phase of the input signal is fixed externally, and uniquely determines all other phases (there is no global 
phase invariance); thus there are 
$2N$ expansion coefficients, (the real and imaginary parts of $a_n^{(\alpha)}$), that must be found, and
an equal number of conditions determining them.

Together Eqs.\ (\ref{IS1}-\ref{Vdef}) define I-SALT.  In the regime in which the SIA holds,
they provide essentially exact solutions of the full coupled wave equations for amplification and injection 
locking. The method is {\it ab initio}, as in SALT, with no prior assumptions about the number, spatial form or frequencies of
the lasing modes. Lasing modes correspond to poles of the non-linear scattering matrix on the real axis; amplified inputs
do not, they are simply additional scattered waves which also deplete the gain.  If the input signal becomes too strong, 
and is sufficiently near in frequency to the lasing mode, then the lasing mode has insufficient gain and
falls below threshold, leaving only the amplified signal output.  The output is ``locked" to the
input frequency, but not by pulling the lasing mode over to $\omega_\alpha$, but rather by turning it off.

\section{From I-SALT to Adler's Model}

In this section we will show how I-SALT can recover an improved version of the traditional Adler equations
in their steady state form. Because our approach starts from the full laser equations we use as comparison
Eqs.\ (58-59) from Spencer and Lamb \cite{lamb72}, which has a similar starting point 
(i.e.\ starts with the full Maxwell equations and includes the
spatial degrees of freedom and gain saturation explicitly). The Adler theory assumes only a single 
input channel with small amplitude and only, at most, a single free-running mode and a single amplified mode;
we model the injected laser following
Spencer and Lamb via a cavity with a perfect mirror at one end and a high reflectivity mirror at the other.
I-SALT is a steady-state theory and should only approach the Adler description 
in the locked regime, thus we assume only a single, highly amplified mode is present in the cavity.
This allows us to approximate the field inside the cavity as
only having two components, one incoming TCF and one outgoing TCF (instead of the full expansion
in outgoing TCFs),
\begin{equation}
\Psi_{in}(x) = au(x; \omega_{in}) + bv(x; \omega_{in}),
\end{equation}
with $a \gg b$ and where $\omega_{in}$ is the frequency of the incident signal.
In a single channel cavity the use of a single incoming TCF is always
justified and in a high-Q cavity the use of a single outgoing TCF is justified by the
Single Pole Approximation (SPA) \cite{ge10}, as the amplified signal is close to a high-Q cavity
resonance and thus only a single outgoing TCF is needed to describe the amplified mode in this limit.
We can use Eqs.\ (\ref{outtcf}) and (\ref{intcf}) to rewrite Eq.\ (\ref{salt_field}) as
\begin{align}
\frac{\gamma_\perp D(x)}{\omega_{in} - \omega_a + i \gamma_\perp} \Psi_{in}(x) &= a \eta_{in} u(x) + b \beta_{in} v(x), \label{spaisalt1} \\
D(x) &=  \frac{D_0}{1 + \Gamma_{in} |\Psi_{in}(x)|^2}. \label{spaisalt2}
\end{align}
By adding and subtracting $b \eta_{in} v(x)$ from the right side of the equation
and defining $\Psi_{in}(x) \equiv a \psi(x)$, we are able to write
\begin{equation}
\frac{\gamma_\perp D(x)}{\omega_{in} - \omega_a + i \gamma_\perp} a \psi(x) - a \eta_{in} \psi(x) = b(\beta_{in} - \eta_{in})v(x).
\end{equation}
While Spencer and Lamb used a delta-function index jump to represent the imperfect mirror,
for convenience we will take our cavity to have a uniform index with the index step at one
end to vacuum comprising the mirror; hence the TCF states will be sine functions
of a complex argument. Thus, in this section only, we choose to normalize our incoming
and outgoing TCF states for convenience as, $(2/L) \int dx u(x) u(x) = 1$, and similarly
$(2/L)\int dx v(x) v(x) = 1$. Integrating through with respect to the mode describing the resonance of the
cavity, $(2/L) \int dx u(x)$, we define the gain saturation function as
\begin{equation}
f(I) = \frac{\omega_{in} D_0}{2 \varepsilon_c} \left(\frac{2}{L}\right) \int dx \frac{u(x) \psi(x)}{1 + \Gamma_{in} I |\psi(x)|^2},
\end{equation}
where $I \simeq |a|^2$ is a measure of the intensity of the field inside of the cavity, and 
has essentially the same meaning as the similar quantity introduced in
Spencer and Lamb, (in their case they use a sine of real argument and are able to evaluate
the resulting integral analytically) \cite{lamb72}. Although $f(I)$ is complex in general, for
high Q cavities, it is essentially real to $10^{-3}$, and thus here we will approximate it as such. Next, we note
that up to corrections of order $b/a$, $(2/L) \int u(x) \psi(x) dx = 1$ as the field profile inside of the
cavity is dominated by the outgoing portion. Finally, the overlap integral between the incoming and
outgoing TCF states can be evaluated by use of Eq.\ (\ref{Vdef}) above, resulting in
\begin{equation}
\frac{\gamma_\perp f(I)}{\omega_{in} - \omega_a + i \gamma_\perp}\left(\frac{2 \varepsilon_c}{\omega_{in}} \right) a - \eta_{in} a = \frac{2 i c}{\omega_{in}}\left(\frac{2}{L}\right) u(0) B_{in},
\end{equation}
where the mirrored side of the cavity has been placed at $x = -L$ and the open edge of the cavity at $x = 0$,
and noting that the definition of the input signal amplitude, Eq.\ (\ref{inputdef}), can be used
to simplify $B_{in} = b v(0)$.
The outgoing TCF eigenvalue for a dielectric slab cavity can be expressed in terms of the input
frequency and the cavity resonance, following Ge \textit{et al.}\ \cite{ge10}
\begin{align}
\eta_{in} =& \varepsilon_c \left( \frac{(\omega_0 - i\frac{\gamma_c}{2})^2 - \omega_{in}^2}{\omega_{in}^2} \right) \notag \\
\simeq& \varepsilon_c \left(\frac{2(\omega_0 - \omega_{in})}{\omega_{in}} - \frac{2i}{\omega_{in}}\left(\frac{\gamma_c}{2}\right) \right),
\end{align}
where $\omega_0$ is the frequency of the passive cavity resonance, $\gamma_c$ is the photon decay rate through the
end of the cavity and
$\omega_0 \simeq \omega_{in}$, as the resonance corresponds to the closest passive cavity resonance
to the injected frequency, at most half a free spectral range away. Finally, we
approximate $(\omega_{in} - \omega_a)^2 \simeq 0$, resulting in
\begin{equation}
\left(\xi - i\right) f(I) a - \left(\Delta - i\frac{\gamma_c}{2} \right)a = \frac{2ic}{L\varepsilon_c} u(0) B_{in},
\end{equation}
where $\xi = (\omega_{in} - \omega_a)/\gamma_\perp$ and $\Delta = \omega_0 - \omega_{in}$.
In high-Q cavities with the normalization for the outgoing TCFs chosen in this section, it
can be shown that $u(0) \simeq 1$. The cavity decay rate can
also be related to the round trip time in the cavity and the reflection coefficient \cite{siegman},
\begin{equation}
\gamma_c = \frac{-c}{2 L n} \ln R,
\end{equation}
where $n$ is the index of refraction of the passive cavity.
Finally, to connect to the Spencer and Lamb version of the Adler theory, we formally 
expand the reflection coefficient for large index, approximating the 
coefficient, $-\ln R \approx T \approx 4/n$, resulting in
\begin{equation}
\gamma_c \approx \frac{2c}{L \varepsilon_c}.
\end{equation}
Finally, writing $B_{in} = |B_{in}|e^{i \phi}$ and separating real
and imaginary components we find
\begin{align}
0 =& \left(f(I) - \frac{\gamma_c}{2} \right) a - \gamma_c |B_{in}| \cos(\phi) \label{IadlerA} \\
0 =& \xi f(I) - \Delta - \gamma_c \frac{|B_{in}|}{a} \sin(\phi). \label{IadlerB}
\end{align}
Noting that in the locked regime $\omega = \omega_{in}$, Eqs.\ (\ref{IadlerA}-\ref{IadlerB}) are identical to
the steady state Adler equations as presented in Eqs.\ (58-59) from Spencer and Lamb \cite{lamb72}, except
that our definition of $f(I)$ includes the openness of the cavity through different boundary condition on
the TCF state, leading to a sine function of complex argument, instead of the Dirichlet boundary condition
assumed in \cite{lamb72}.

Thus, in the correct limit, an improved version of the traditional theory can be recovered from I-SALT, in
the locked regime.  If we make the further usual assumption that $\xi$ is small, we will obtain exactly the
same locking range as predicted by the standard theory (see next section).
In the unlocked regime the Adler theory predicts a residual time-dependence of the
relative phase of input and free-running signal which cannot be derived from I-SALT; 
but full I-SALT shows that the frequency shifts predicted by the usual theory in the unlocked
regime are not correct in general (next section).

\section{Numerical Results}

To test the results of I-SALT, we compare them to the exact numerical solutions from 
Finite Difference Time Domain (FDTD) simulations of the Maxwell-Bloch equations for 
a simple one-dimensional asymmetric Fabry-Perot cavity with an injected signal (schematics) under various conditions.
The FDTD simulations performed here used the time-stepping method proposed by Bid\'{e}garay, updating the atomic polarization and inversion
alongside the magnetic field, and were run for a total time of $T_{tot} \sim 100 (1/\gamma_\parallel)$ to ensure convergence, as
$\gamma_\parallel$ corresponds to the longest time-scale in the system \cite{bidegaray03}.  Similar simulations without 
the injected signal were previously used for quantitative tests of SALT \cite{ge08,cerjan12}.

\subsection{locking transition}
\begin{figure}
\centering
\includegraphics[width=0.45\textwidth]{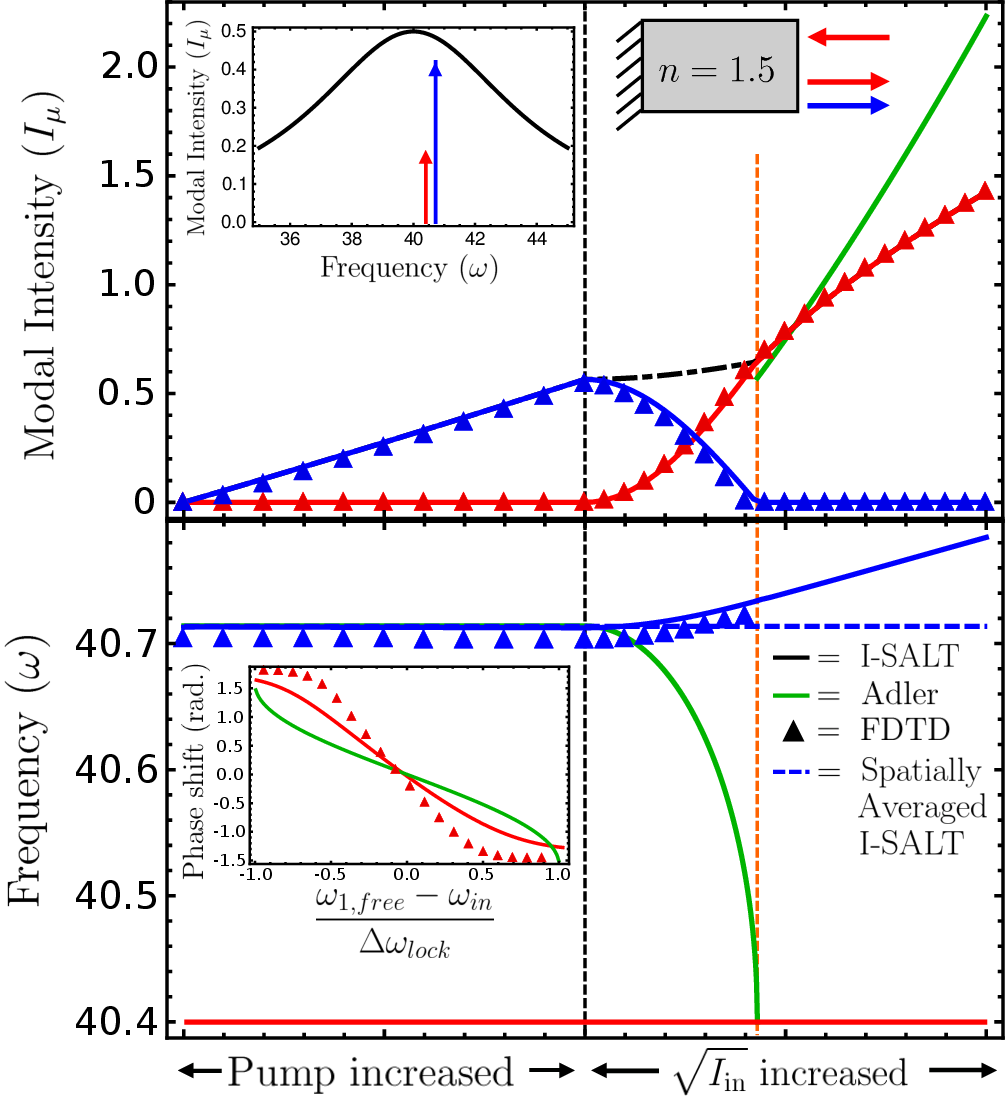}
\caption{(Top panel) Simulations of single mode injection locking in 
  a one-sided dielectric slab cavity with $n=1.5$ (schematic) with a perfect mirror
  at one end and index step to vacuum at the other. First the pump
  is increased above the threshold for lasing at $\omega_{1,free} = 40.714$, $D_0 = 0.0603$ to $D_0 = 0.08$, and then held at a fixed value (vertical
  black line) while the input
  signal amplitude is ramped from $B_{in} = 0$ until the free-running signal is quenched and the system ``locks" (vertical orange line)
  to the injected frequency, $\omega_{in} = 40.4$ at $B_{in} = 0.176$. Finally, the simulation is continued in the locked regime to $B_{in} = 0.4$.
  Solid lines are output intensities calculated from I-SALT; blue is lasing output, 
  red is amplified output at signal frequency, dot-dashed black is total output. Triangles are 
  the same quantities from FDTD for the same dielectric slab laser with $\omega_a = 40$, 
  the width of the gain curve, $\gamma_\perp = 4$, and $\gamma_\parallel = 0.001$.  Green curve in the locked regime is the prediction of our
  generalized Adler equations, (\ref{IadlerA}-\ref{IadlerB}).  Top inset shows gain curve and $\omega_{in}$ (red), $\omega_1$
  (blue). (Bottom panel) Frequency variation of the first lasing mode:
  blue line is from I-SALT and blue triangles from FDTD. The green line shows the prediction of the Adler theory.
  The red line is the injected signal frequency. Again, the orange 
  dashed line shows the locking threshold from I-SALT, frequencies beyond this point are taken as
  the real part of the location of the pole of the scattering matrix. As the locking transition is approached 
  the lasing frequency moves {\it away} from the injected frequency, due to spatial hole-burning instead of
  being attracted toward it as expected from the Adler equation \cite{siegman}. 
  Blue dashed lines showing negligible frequency shift are I-SALT calculation with uniform gain saturation
  and no spatial hole-burning.
  The inset shows a plot of the phase shift between input and output signals of an injection locked
  dielectric slab cavity at a fixed input intensity. I-SALT (red) and FDTD simulations (red triangles)
  are seen to have a better quantitative and qualitative agreement than the Adler prediction (green). For comparison
  with the Adler theory, the horizontal axis is plotted in terms of the free-running lasing frequency in the 
  absence of an injected signal at threshold, $\omega_{1,free}$. Values quoted
  are given in units of $c/L$. \label{s_l}}
\vspace{-10pt}
\end{figure}

We first study the usual locking transition in Fig.\ \ref{s_l} in which a single free-running mode eventually gives
way to an injected mode.  The simulations are done in a region of large detuning in which we 
expect good agreement with I-SALT. Indeed locking of the output signal
to the input is found in the FDTD data in good quantitative agreement with I-SALT with no adjustable
parameters. We note that the quantitative agreement seen between I-SALT and FDTD calculations is also a demonstration of 
the stability of the I-SALT solutions; any instabilities due to beating terms in the inversion would be present in the FDTD solutions, which do not rely on the stationary inversion approximation.
A further analytical treatment of the stability of the I-SALT solutions will be presented in section VI below.
The simulations in Fig.\ \ref{s_l} are for large detuning, not the typical Adler regime; thus for
these parameters locking requires an input signal which is a significant fraction of the free-running output
at that pump value, $\sim 23 \%$.  The total output intensity of the amplified mode when locking occurs is larger 
than the free running signal also by $\sim 18 \%$ (in the Adler theory they are the same to a good approximation), 
however this is not surprising due to the relatively large input intensity.  Also the independence of the spatial degrees of freedom
of the amplified and free-running mode should allow the amplifier to extract more power from the gain medium.  
Consistent with this, in the unlocked region, when both free-running and amplified modes are emitting, the total 
output intensity is monotonically increasing, as indicated by the black dashed line in the figure.
If we take the I-SALT version of the Adler equations in the locked regime, and impose the condition that locking occurs when
the amplified output is equal to the original free-running output, we can solve the non-linear Eqs.\ (\ref{IadlerA}-\ref{IadlerB}) to predict
the input amplitude at which the laser would lock. This transition line is very close to that found by I-SALT, slightly less by $\sim 4.5 \%$. 
Above that point we can plot the Adler I-SALT predictions for the amplified mode intensity and find them to be in reasonable agreement with I-SALT and FDTD near
the transition and in poor agreement far above it.

The difference between the locking behavior in this regime and in the usual Adler theory is strikingly illustrated
by the frequency shift of the free-running mode prior to locking, which is qualitatively different from the Adler theory; the 
free-running frequency, $\omega_{1}$, is repelled from the input frequency (full blue line and data points), instead of being strongly attracted toward it (green line).  
This frequency repulsion can be explained by the combined effects of mode competition and spatial hole-burning. As the incident signal is imposed and 
depletes the gain, the standing wave of the laser field shifts away from the frequency/wavelength of the 
incident standing wave in order to better extract energy from the regions of the cavity where the gain is not being
saturated by the incident signal. To confirm this interpretation we replaced the {\it space dependent} gain saturation      
denominator with its spatial average,
\begin{equation}
D_{ave}(x) = \frac{d_0}{1 + \sum_\sigma^{N_L+N_A} \frac{\Gamma_\sigma}{L} \int_C |\Psi_\sigma(x)|^2 dx},
\end{equation}
in place of Eq.\ \ref{ampinv}, and noting that for this simulation the pump profile is uniform, $F(x) = 1$.
Using this spatially averaged gain saturation, no movement of the lasing frequency is seen in Fig.\ \ref{s_l}b (dashed blue line).
This provides strong evidence that the frequency pushing phenomenon observed here requires treating the full spatial dependence
of the problem and can not be seen in previously developed spatially averaged injection theories, \cite{oppo_pushing_86,solari_94,zimmermann_global_01}.

Furthermore, as I-SALT and FDTD simulations both predict this same frequency repulsion and this solution is found to be
stable (see Section VI), the effect seen here is different from the frequency repulsion 
previously predicted in dynamical parameter regimes by Oppo \textit{et al.}, where the SIA would
not be appropriate and as a result more complex dynamical features are found \cite{oppo_pushing_86}.
The relatively weak repulsion 
found here is also distinct from that observed by Murakami \textit{et al.}\ \cite{murakami03}, which is a shift
in the cavity resonance due to the injected signal saturating the gain carriers in materials with large
Henry $\alpha$-factors such as semiconductors where $\alpha \sim 2-8$. In contrast, for the Bloch gain
medium used in Fig.\ \ref{s_l}, $\alpha = 0.17$. Additionally, the overall saturation of the gain medium
is not changing significantly while this frequency repulsion is observed as the pump is held fixed
while the injected signal is increased, thus keeping the total output intensity relatively constant, as is
seen in Fig.\ \ref{s_l}a in the black dot-dashed line. For the effect predicted by Murakami \textit{et al.}\
to be seen, a significant shift in the number of available gain carriers is needed, coupled to a large
Henry $\alpha$-factor, and this effect would be seen in the spatially-averaged I-SALT calculation if it were present \cite{murakami03}.

As noted, the Adler theory describes locking driven by phase synchronization of the input and free-running
fields.  Since the threshold input intensity for locking decreases to zero as the input frequency, $\omega_{in}$, approaches
the free-running frequency in the absence of an injected signal, $\omega_{1,free}$, the threshold condition can be expressed as a ``locking range", the frequency range
$\Delta \omega_{lock}$, over which the laser is locked for a given input intensity.  In the Adler theory one finds
\begin{equation}
  \Delta \omega_{lock} = \gamma_c \sqrt{\frac{|B_{in}|^2}{I_{0}}}, 
\end{equation}
where $\gamma_c$ is the cavity decay rate, $|B_{in}|^2$ is the intensity of the input signal, and $I_{0}$ is
the intensity of the free running lasing signal in the absence of the input \cite{siegman}.
Within this locking range there is a fixed phase relationship between the input signal and the locked output
which varies as 
\begin{equation}
  \Delta \phi = \arcsin \left[\frac{\omega_{1,free} - \omega_{in}}{\Delta \omega_{lock}} \right].
\end{equation}
The same quantity can be calculated in I-SALT and is compared to the Adler prediction for the same slab cavity in 
the inset of Fig.\ \ref{s_l}a.
The phase shift variation found from I-SALT is substantially different from the Adler prediction and in
good agreement with FDTD (see inset, Fig.\ \ref{s_l}b).

\begin{figure}
\centering
\includegraphics[width=0.45\textwidth]{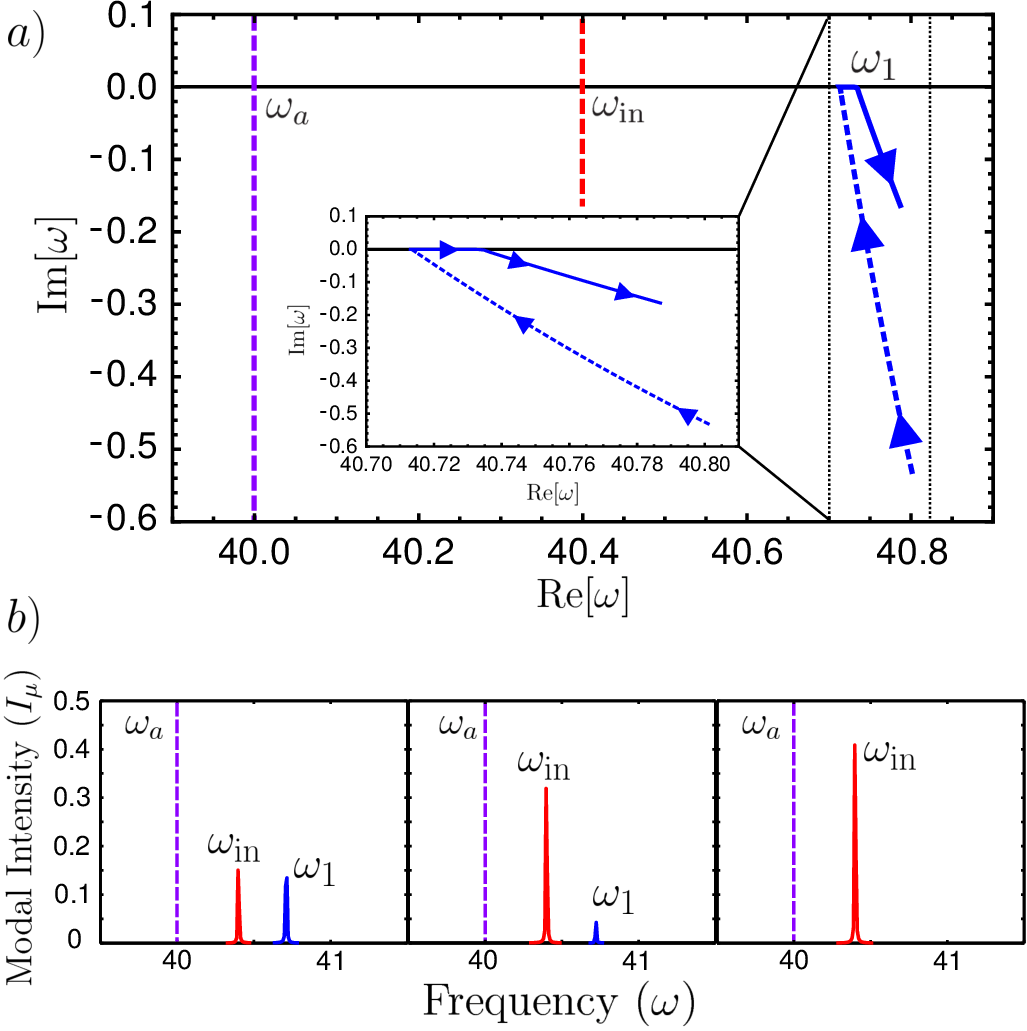}
\caption{ (Top panel, a) Motion of the pole as described in the text corresponding to the free-running lasing mode in the locking scenario of Fig.\ \ref{s_l}.
  As the pump is increased below threshold the pole of the scattering matrix 
  is pulled upwards towards the real axis and ``in" towards $\omega_a=40$ (blue dashed line, recall there is no signal yet at 
  $\omega_{in} = 40.4$).
  Free-running lasing occurs when the pole reaches the real axis at $\omega_{1,free} = 40.714$ and continues
  as the pump is increased further above threshold with negligible further frequency shift.
  Then the pump is fixed and the input signal is ramped,
  causing the pole (solid blue line) to move to higher frequency, away from the input frequency, and eventually off the real axis as the effects of gain saturation cause the lasing mode to go below threshold.
  Inset shows a zoom-in on the motion of the pole of the
  lasing mode, inside of the dotted box.
  (Bottom panel, b) Frequency spectrum from FDTD simulations across the locking transition, showing no 
  additional lines appearing, indicating that the effect is purely due to gain cross-saturation.
\label{poleplot}}
\end{figure}

The fact that in this regime the locking transition is entirely due to gain cross-saturation, with no contribution
from beating or phase synchronization, is illustrated in Fig.\ \ref{poleplot}a. 
In the top panel we show the motion of the pole of the scattering matrix
corresponding to the lasing mode in Fig.\ \ref{s_l} as the pump is increased and then fixed, and then the signal
is injected and increased.  The dashed blue line corresponds to the laser being below threshold; as the pole moves up
towards the real axis its real frequency, $\omega_1$, decreases, pulled toward the center of the atomic line, $\omega_a$.  When the pole
reaches the real axis, corresponding to the free-running threshold, $\omega_{1,free}$, the gain balances loss and the mode lases.
As the pump is further increased, the pole moves slightly further toward the
center of the gain curve (not visible on this scale), but as soon as the pump is fixed and the injected signal is
turned on at $\omega_{in} < \omega_1$, the behavior reverses. As the injected signal increases the lasing frequency increases, 
shifting away from $\omega_{in}$ (and $\omega_a$). Eventually the injected mode saturates the gain enough to
drive the lasing mode below threshold, and the pole leaves the real axis, although it continues to be repelled
from $\omega_{in}$.  This demonstrates that in the regime of stationary atomic populations the locking transition
corresponds simply to driving the lasing mode below threshold due to the saturation of the gain medium from the
injected mode. 

A final important indication of the non-Adler nature of the transition is given in Fig.\ \ref{poleplot}b.  In the Adler theory
there are always strong four-wave mixing effects as the locking threshold is reached \cite{siegman}, and additional lines
should appear in the frequency spectrum.  In Fig.\ \ref{poleplot}b we show the Fourier transforms of the FDTD data across the 
locking transition, which indicated a smooth transfer of intensity from the free-running line to the injected line with
no additional frequencies appearing as the free-running line disappears.

\begin{figure}
\centering
\includegraphics[width=0.45\textwidth]{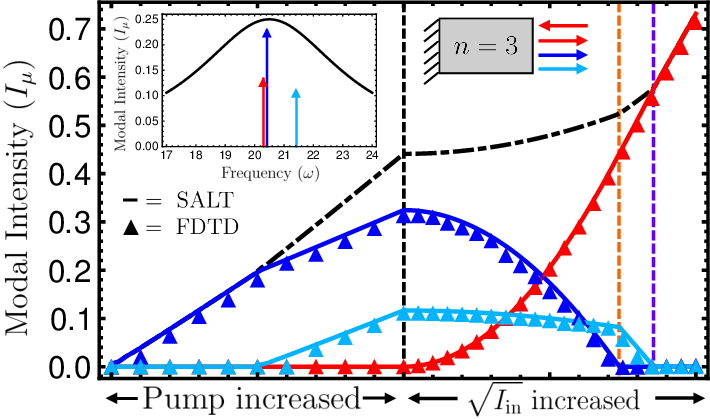}
\caption{Partial locking transition as described in the text for a laser with two free-running modes and
  an injected signal (schematic) using a similar pumping
  and input ramping scheme as Fig.\ \ref{s_l}, starting at the first lasing threshold $D_0 = 0.101$ and pumping until $D_0 = 0.13$,
  then increasing the input signal from $B_{in} = 0$ to $B_{in} = 0.4$.
  Solid lines are output intensities calculated from I-SALT; blue and cyan are lasing output, 
  red is amplified output at signal frequency, $\omega_{in} = 20.3$, dot-dashed black is total output. Triangles are 
  the same quantities from FDTD for a similar dielectric slab laser with $n=3$, $\omega_a = 20.5, \gamma_\perp = 3, \gamma_\parallel = 0.001$ Inset shows the relationship of the three frequencies. 
  As expected, the lasing mode nearest to the injected signal locks to the injected signal (orange line),
  then the more distant lasing mode locks (purple line). Values quoted are given in units of $c/L$.
  \label{p_l}}
\end{figure}

\subsection{partially locked states}

Beyond yielding a correct and quantitative treatment of the locking transition of a single mode laser in the relevant
regime, as seen from the generality of Eqs.\ (\ref{IS1}-\ref{Vdef}), I-SALT is able to treat simultaneously multimode lasing
with multiple inputs.  An interesting example is shown in Fig.\ \ref{p_l}. Here an asymmetric Fabry-Perot slab laser similar to 
that in studied in Fig.\ \ref{s_l} is pumped above its {\it second} lasing threshold and a signal is injected closer to the frequency of the first lasing mode.  
Because of its stronger interaction with the first mode (blue), the
signal is able to lock that mode, while the second mode (cyan) remains active at a similar frequency to its free-running value, 
though shifted away from the injected frequency in the same manner as described before.
As before the solid lines (I-SALT) are in good agreement with the data points (FDTD).
Thus, with relatively little additional computational effort, I-SALT predicts such ``partially-locked" states, something which is not treated in previous theories.  

\begin{figure}
\centering
\includegraphics[width=0.45\textwidth]{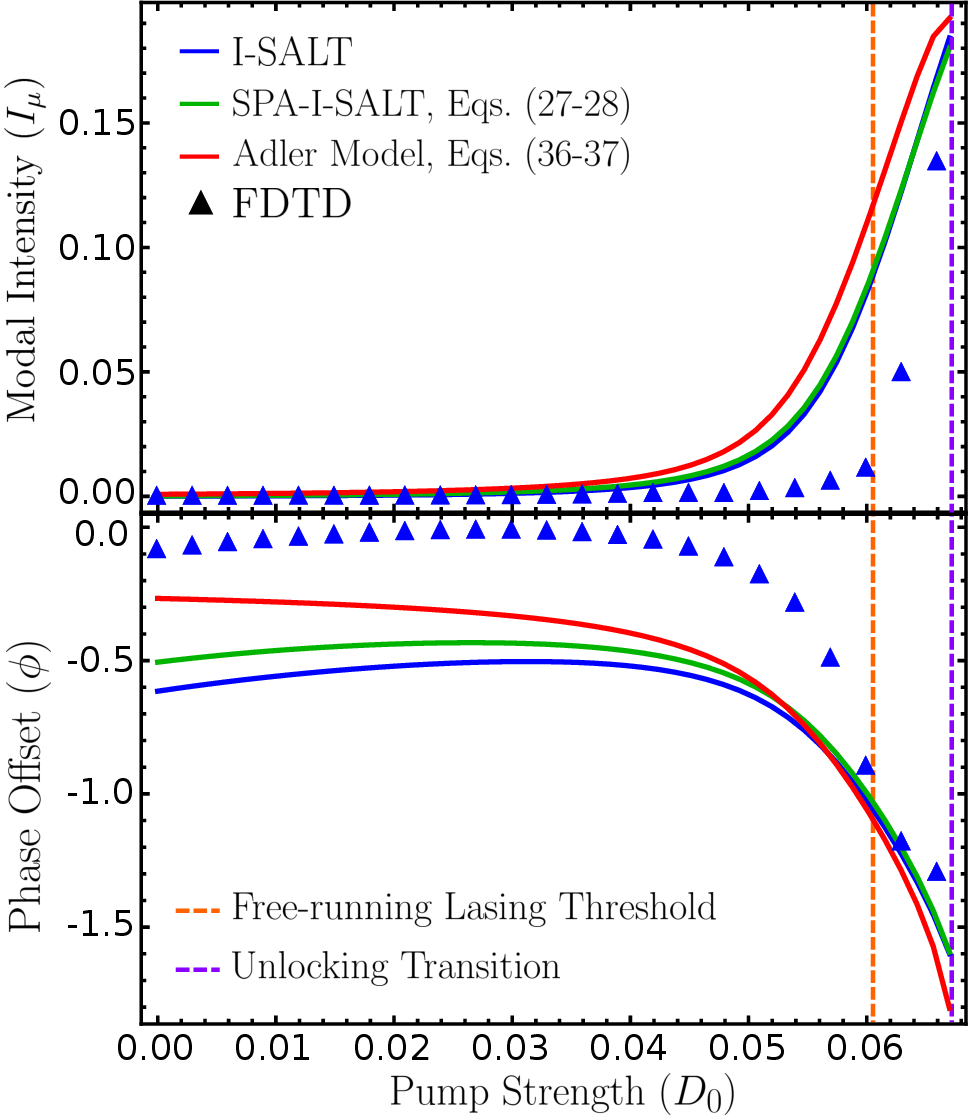}
\caption{Comparison of the predictions of I-SALT, blue line, the single pole approximation of I-SALT,
  given by Eqs.\ (\ref{spaisalt1}-\ref{spaisalt2}), green line, the Adler model, as given by Eqs.\
  (\ref{IadlerA}-\ref{IadlerB}), red line, and FDTD, blue triangles for both the output intensity (top)
  and the phase offset (bottom).  The input intensity is negligible
  ($|B_{in}|^2 = 10^{-4}$) compared to the output, while the pump (gain) is increased from $D_0 = 0$ to $D_0 = 0.067$, thus placing the simulations in
  the regime of validity for the Adler approximations. The vertical orange dashed line denotes
  the first lasing threshold in the absence of an incident signal, whereas the vertical purple
  dashed line shows where I-SALT predicts the unlocking transition to occur.
  Simulations are shown for a single-sided dielectric slab
  cavity with $n=1.5$, $\omega_a = 40$, $\omega_{in} = 40.7$, $\omega_{1,free} = 40.714$, $\gamma_\perp = 4$,
  and $\gamma_\parallel = 0.001$, units quoted in values of $c/L$.
  \label{amplifier}}
\end{figure}

\subsection{I-SALT and Adler I-SALT below threshold}

The previous results assumed rather large detuning and hence relatively large injected signals to reach locking.
To test our work in the more conventional regime of small detuning and small injected signals we consider 
injection near the free-running lasing frequency as a function of pump.  Here we are near the center of the gain spectrum
and will have much higher amplification.  Since I-SALT is not reliable in the unlocked regime for detuning smaller than the
relaxation oscillation frequency, we study only the behavior for pump values below the lasing threshold, when the cavity is 
functioning as a regenerative amplifier for the injected signal.  Since there is only emission at the injected signal in this regime, 
we can also apply the Adler approximation to I-SALT.  As shown in Fig.\ \ref{amplifier}, even though I-SALT may not describe 
well the unlocked regime for this cavity, it provides a very accurate description of the amplifier, in good agreement with FDTD 
for both intensity and phase offset.  We also find, as one might expect, that the Adler approximation to I-SALT (red curves) 
works almost as well. The dashed vertical lines in the figure show the lasing thresholds in the absence of the injected signal 
(orange) and in its presence (purple); note that the injected signal pushes up the lasing threshold significantly.

\subsection{Injected Quadrupole Resonators \label{sec5b}}

As noted in our introduction, a strength of the SALT and I-SALT theories is that they can handle
an arbitrary cavity geometry essentially exactly.  Here we demonstrate the power of the method
by simulating injected two-dimensional quadrupole resonators, below the first lasing threshold. The boundary of the quadrupole cavity is defined by
\begin{equation}
R(\phi) = R_0(1 + \epsilon \cos(2\phi)),
\end{equation}
where $\phi$ is the polar angle, $R_0$ is the average radius, and $\epsilon$ is the deformation parameter.
Disk or cylinder resonators of this type have been of interest for some time \cite{nockel97,gmachl98}
because as a function of the deformation the ray dynamics in the cavity undergoes a transition to chaos,
with an attendant change in the emission patterns from the laser modes.
For thin disks in three dimensions strictly speaking one should treat the diffraction effects in the 
axial ($z$) direction; for cylinders many wavelengths long one may treat them as infinite in the $z$-direction
and study the $k_z=0$ mode, which then reduces to this purely two-dimensional scalar problem for 
either the electric (TM) or magnetic (TE) modes.  It is slightly simpler to treat the TM case for
which the electric field is continuous at the boundary and we will focus on that case here. 
Both SALT and I-SALT are both capable of treating
modes of arbitrary polarization \cite{esterhazy14}. In two dimensions, the boundary condition
for the incoming and outgoing TCF states requires matching via continuity from the interior cavity solutions
to exterior solutions consisting of a superposition of either incoming or outgoing Hankel functions.
The detailed method for doing this has been previously described \cite{tureci08,li_thesis},
and for brevity we will omit it here. As mentioned above, the injection profile must
now be defined at the border of the entire two dimensional cavity, which must then be matched
to an expansion of the incoming TCF states along this boundary.

\begin{figure}
\centering
\includegraphics[width=0.45\textwidth]{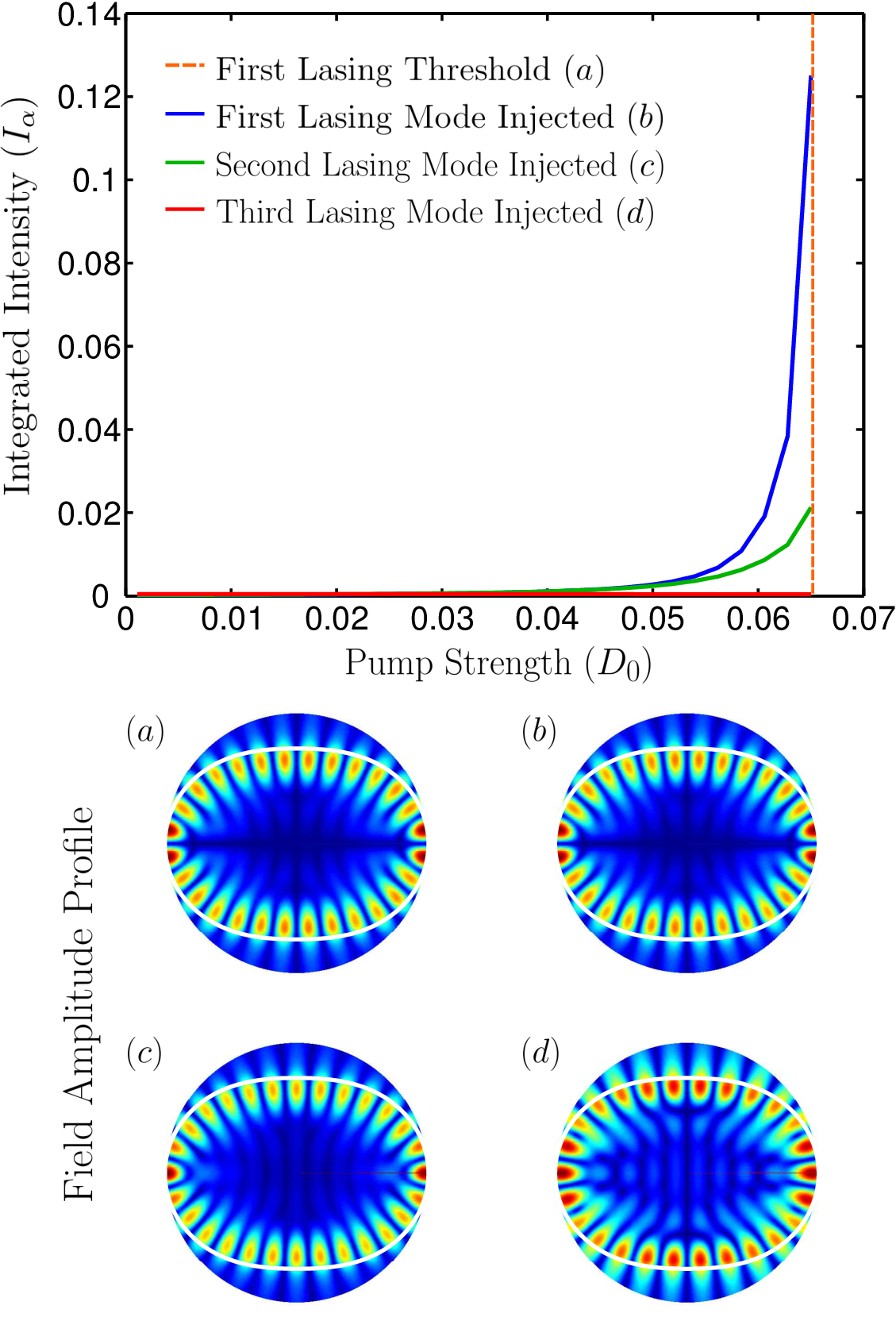}
\caption{Simulations of a uniform index quadrupole cavity laser amplifier with $n=1.5$ (boundary 
indicated in white). The parameters chosen are $R_0 = 1.72 \mu m$,
  $\lambda_a = 1\mu m$, $\gamma_\perp = 0.03 \mu m$, and $\epsilon = 0.16$. The injected
  wavelength for all three simulations is the same as the that for the first free-running
  mode, $\lambda_{in} = \lambda_1 = 0.989\mu m$. The pump value was increased from $D_0 = 0$ to the first
  free-running lasing threshold, $D_0 = 0.065$, vertical orange dashed line. The three solid curves show the amplifier
  output for three different injection conditions: blue, injection with the TCF corresponding to the 
  first lasing mode; green, injection with the TCF corresponding to the second lasing mode; red,
  injection with the TCF corresponding to the third lasing mode.
  Lower plots show in color scale the normalized mode amplitude profiles:
  (a) The first free-running lasing mode at threshold. (b) The amplified mode with the first lasing mode's incoming
  TCF as input. (c) The amplified mode with the second lasing mode's incoming TCF as input. (d)
  The amplified mode with the third lasing mode's incoming TCF as input. The full disk shown in blue is the simulation region used; only TM modes were simulated.\label{fig5}}
\end{figure}

Unlike the simple injection boundary condition in a single dimension, Eq.\ (\ref{inputdef}),
in two dimensions there is an infinite variety of injected fields at the boundary and 
we find that the choice of the injection profile plays a large role in determining
the resulting amplified mode profile, as seen in Fig.\ \ref{fig5}. Here, the cavity is injected
with three different injection profiles, all at the first lasing mode's free-running frequency. The first lasing
threshold in the absence of injection is shown as the orange dashed line and its corresponding
mode profile is shown below. When the injected signal is given 
by the incoming TCF corresponding to the first lasing mode at threshold, the injected signal
is amplified dramatically, and the resulting mode profile is nearly identical to that of the
threshold lasing mode. However, when the injected signal, still at the same frequency, is chosen to be
the incoming TCF corresponding to the second lasing mode, substantially less
amplification occurs and the resulting mode profile is very similar to that of the second lasing mode at
threshold, even though the difference in frequency between the first and second lasing modes
is small compared with the linewidth of the atomic transition, and they have similar thresholds.
Finally, when the incoming signal is represented by the incoming TCF state corresponding to the
third lasing mode, almost no amplification occurs, and the mode profile does not resemble the third
lasing mode. Evidently, the injected mode profile plays a much stronger role in choosing the resulting
amplified mode than the injected frequency does. Essentially the injected profile is playing the role of a coherent 
pump or seed with a strong selectivity for a given resonance, so that we may think of it as interchanging the thresholds
for e.g. the first and second modes.  This qualitative conclusion may have been difficult to guess in the 
absence of a quantitative theory.

\section{Stability Analysis}

The excellent agreement found between I-SALT and FDTD simulations in the previous section is a
good indication that the solutions of I-SALT are stable in time. To confirm this fact,
we now perform a stability analysis of the I-SALT solutions under two simplifying
assumptions. First, we will continue to make the assumption from SALT that the beating terms
of the form $\exp[-i(\omega_\sigma - \omega_\nu)t]$ time average to zero, where $\omega_\sigma$ and $\omega_\nu$ are the
frequencies of lasing or amplified modes in the system. Second, we will only consider
spatially uniform perturbations to the lasing and amplified modes. To address instabilities originating 
from the beating terms in the inversion equation that are neglected in I-SALT, a different
analysis can be performed which is nearly identical to that presented by Ge \textit{et al.}\ in which
the magnitude of the beating the atomic inversion can be calculated \cite{ge08}. To treat the
more general problem of spatially dependent perturbations a more detailed analysis is being performed by
Stefan Rotter and Dmitry Krimer \cite{rotter_stability}. However, in all previous treatments of stability
for injected systems of which we are aware, the slowly varying envelope approximation (SVEA) is
invoked, eliminating the spatial degrees of freedom for those treatments as well.

Starting from the Maxwell-Bloch equations, we again insert a modal decomposition of
the electric field and polarization, where the amplitudes have been decomposed into
their steady state values, $\bar E, \bar P$, and the time dependent perturbations, $\delta E, \delta P$,
\begin{align}
  E^+ &= \sum_\sigma  (\bar E_\sigma + \delta E_\sigma(t)) \Psi_\sigma(x) e^{-i \omega_\sigma t} \\
  P^+ &= \sum_\sigma  (\bar P_\sigma + \delta P_\sigma(t)) p_\sigma(x)e^{-i \omega_\sigma t}.
\end{align}
The inversion can also be decomposed in a similar manner, but with only the slowly-varying part and
no ``carrier frequency", 

\begin{equation}
  D(x,t) = (\bar D + \delta D(t) )d(x). 
\end{equation}
These expansions are then inserted back into the Maxwell-Bloch equations in which we are assuming
the cavity dielectric is a constant, the steady state
behavior is removed, and second derivatives of the perturbations are assumed to be much smaller
than the other terms, to find
\begin{align}
  4 \pi & \left(-\omega_\sigma^2 \delta P_\sigma -2i \omega_\sigma \dot{\delta P}_\sigma \right) p_\sigma(x) = \notag \\
  &\delta E_\sigma (\nabla^2 + \varepsilon_c \omega_\sigma^2) \Psi_\sigma(x) + 2i \varepsilon_c \omega_\sigma \dot{\delta E}_\sigma \Psi_\sigma(x) \label{stab2a}
\end{align}
\begin{align}
  \dot{\delta P}_\sigma p_\sigma(x) =& (i\omega_\sigma -i \omega_a - \gamma_\perp)\delta P_\sigma p_\sigma(x) \notag \\
  &+ \frac{\gamma_\perp}{4 \pi i} (\delta D E_\sigma + D \delta E_\sigma) d(x) \Psi_\sigma(x) \label{stab2b}
\end{align}
\begin{align}
  \dot{\delta D} d(x) =& -\gamma_\parallel \delta D d(x)\notag \\
  &+ 2 \pi i \gamma_\parallel \left( \sum_\sigma (\bar E_\sigma \delta P_\sigma^* + \bar P_\sigma^* \delta E_\sigma)\Psi_\sigma(x) p_\sigma^*(x) \right. \notag \\ 
  &\left. - c.c. \right). \label{stab2c}
\end{align}
The linearized stability equations (\ref{stab2a}-\ref{stab2c}) can be further simplified through the
use of the known steady state solutions,
\begin{align}
  - 4 \pi \omega_\sigma^2 \bar P_\sigma p_\sigma(x) =& \bar E_\sigma (\nabla^2 + \varepsilon_c \omega_\sigma^2) \Psi_\sigma(x) \label{stab3a} \\
  \bar P_\sigma p_\sigma(x) =& \frac{\gamma_\sigma}{4\pi} \bar D \bar E_\sigma d(x) \Psi_\sigma(x) \label{stab3b}
\end{align}
\begin{align}
  0 = \gamma_\parallel(D_0 - \bar D d(x)) + 2 \pi \gamma_\parallel i \left(\bar E_\sigma \bar P_\sigma^* \Psi_\sigma(x) p_\sigma^*(x) - c.c. \right), \label{stab3c}
\end{align}
which allows for the removal of the spatial profiles of the modes. As such,
the evolution of the perturbation of the polarization, (\ref{stab2b}) can be rewritten
by dividing through by $\bar P_\sigma p_\sigma(x)$, and using the steady-state solution (\ref{stab3b}), to find,
\begin{equation}
  \frac{\dot {\delta P}_\sigma}{\bar P_\sigma} = \frac{i \gamma_\perp}{\gamma_\sigma} \left(\frac{\delta P_\sigma}{\bar P_\sigma} - \frac{\delta E_\sigma}{\bar E_\sigma} - \frac{\delta D}{\bar D} \right). \label{stabpol}
\end{equation}

To simplify the perturbations in the wave equation (\ref{stab2a}), one can first evaluate
the derivative of the spatial mode profile through the use of the steady state solution of the
wave equation (\ref{stab3a}). Next, (\ref{stab3b}) is used to rewrite the remaining spatial
dependence in terms of the inversion, and finally we integrate both sides with respect to
$1/V \int_C d^dx$, resulting in
\begin{equation}
  \frac{-2 i \varepsilon_c}{\gamma_\sigma \langle \bar D d(x) \rangle} \left(\frac{\dot{\delta E}_\sigma}{\bar E_\sigma}\right) 
  -2 i \left(\frac{\dot{\delta P}_\sigma}{\bar P_\sigma} \right) = \omega_\sigma \left( \frac{ \delta P_\sigma}{\bar P_\sigma} - \frac{ \delta E_\sigma}{\bar E_\sigma} \right), \label{stabwav}
\end{equation}
where
\begin{equation}
\langle \bar D d(x) \rangle = \frac{1}{V} \bar D \int_C d(x) d^dx
\end{equation}
is the spatial average of the inversion. As we are not considering spatially dependent
perturbations, only global changes in the amplitudes of the fields in the problem, treating
the spatial variation of the inversion would violate our previous assumptions. It should
be noted that this spatial average could also be performed at the outset (\ref{stab2a}-\ref{stab2c}) without changing
any of these results.

Finally, the evolution of the perturbation in the inversion (\ref{stab2c}) can be rewritten
using the steady state of the polarization (\ref{stab3b}) and its complex conjugate, and
again integrating over the cavity, to find
\begin{align}
  \frac{\dot{\delta D}}{\bar D} = -\gamma_\parallel \frac{\delta D}{\bar D} + &\left(\frac{i \gamma_\parallel}{2} \right) \sum_\sigma \langle |\bar E_\sigma \Psi_\sigma(x)|^2 \rangle \notag \\
  &\times\left[ \gamma_\sigma^* \left( \frac{\delta P_\sigma^*}{\bar P_\sigma^*} + \frac{\delta E_\sigma}{\bar E_\sigma} \right) - c.c. \right], \label{stabinv}
\end{align}
where
\begin{equation}
  \langle |\bar E_\sigma \Psi_\sigma(x)|^2 \rangle = \frac{1}{V} \int_C |\bar E_\sigma \Psi_\sigma(x)|^2 d^dx
\end{equation}
is the spatial average of the lasing mode profile.

The evolution equations for the perturbations, (\ref{stabpol}), (\ref{stabwav}), and (\ref{stabinv}) collectively comprise $4(N_L+N_A)+1$
independent equations, four equations per mode for the real and imaginary portions of (\ref{stabpol}) and
(\ref{stabwav}), and a single real equation for the perturbation to the atomic inversion (\ref{stabinv}), which
couples all of the active modes together. The last step which is standard in a stability
analysis calculation is to assume solutions of these equations in the form
\begin{equation}
  \delta E_\sigma, \delta P_\sigma, \delta D \propto e^{\lambda t}
\end{equation}
and ensure that all of the
solutions are decaying, $\textrm{Re}[\lambda] < 0$. However, in the case of a lasing mode there
is an undetermined global phase of the mode, thus for every lasing mode in the calculation
we expect a single marginal eigenvalue, $\textrm{Re}[\lambda] = 0$, corresponding to the lack
of a restoring force for the phase of the lasing mode.
We also expect a single marginal eigenvalue for each amplified mode as well. This is a reflection
of the fact that the amplitude perturbation of an injected mode being considered here is affecting both the incoming
and outgoing portions of the mode equally,
\begin{align}
  \delta E_\alpha(t) \Psi_\alpha(x) = \delta E_\alpha(t)  &\left(\sum_n a_n^{(\alpha)} u_n(x; \omega_\alpha) \right. \notag \\
  &\left. + \sum_m b_m^{(\alpha)} v_m(x; \omega_\alpha)\right),
\end{align}
and thus has the ability to change the global phase of both the incoming and outgoing components
of the mode. However, it must alter both portions by the same phase shift, thus leaving the relative
phase difference between the incident and outgoing components fixed, as is expected.

The results for this spatially averaged stability analysis for the simulations shown in Fig.\ \ref{s_l}
can be seen in Table \ref{tab1}. The first column shows the eigenvalues for the cavity when only
a single lasing mode is active, right before the pump is fixed and the incident mode is turned on,
and the I-SALT solution is found to be stable.
As expected we find four decaying eigenvalues and a single marginal eigenvalue. The second
column shows the eigenvalues for the cavity when both the lasing and amplified modes are present
in the cavity, at nearly the location where their output intensities are equal. Again, the
I-SALT solution found is stable, with only decaying or marginal eigenvalues found, however, an
extra marginal eigenvalue is found, which was not anticipated. The analysis of why there
is an extra marginal eigenvalue is left for the future, more complete stability analysis
of the SALT and I-SALT solutions.

\begin{table}
  \begin{tabular}{c | c | c}
     & $D_0 = 0.075$, $B_{in} = 0$ & $D_0 = 0.08$, $B_{in} = 0.10$ \\
    \hline
    $\lambda_1$ & $-4.60878 + 0.70576i$ & $-4.61428 + 0.71661i$ \\
    \hline
    $\lambda_2$ & $-4.60878 - 0.70576i$ & $-4.61428 - 0.71661i$ \\
    \hline
    $\lambda_3$ & $-0.00088$ & $-0.00084$ \\
    \hline
    $\lambda_4$ & $-0.00012$ & $-0.00016$ \\
    \hline
    $\lambda_5$ & $0$ & $0$ \\
    \hline
    $\lambda_6$ & -- & $-4.60483 + 0.34413i$ \\
    \hline
    $\lambda_7$ & -- & $-4.60483 - 0.34413i$ \\
    \hline
    $\lambda_8$ & -- & $0$ \\
    \hline
    $\lambda_9$ & -- & $0$ \\
  \end{tabular}
  \caption{Stability eigenvalues for the cavity shown in Fig.\ \ref{s_l} at two different
    locations of pump and injection strength. The first column shows the eigenvalues right before
    the injected mode is turned on, with only a single lasing mode active in the cavity. The second
    column shows the eigenvalues when both the lasing mode and amplified mode are present in the
    cavity and have nearly the same output intensity. In both cases, the I-SALT solution is
    found to be stable, though in the second case, an extra marginal eigenvalue is found.\label{tab1}}
\end{table}

\section{Conclusion}

In this work we have developed a new method of treating the steady-state behavior of
laser with injected signals which gives an exact treatment of both the openness of the cavity
and the effects of spatial hole-burning. The theory is valid for Class A and B lasers as long
as $\gamma_\parallel$ and the relaxation oscillation frequency are sufficiently small compared to the other relevant 
frequency scales as discussed in the introduction.  These conditions are typically met in modern microlasers.
I-SALT was then shown to predict qualitatively different
behavior of the locking transition from previous Adler-like theories which describe the process as one of phase
synchronization. When the atomic inversion is stationary, 
the locking transition is entirely due to spatially-varying gain-saturation which has the effect of pushing the
frequency of the lasing mode away from the frequency of the injected mode, in contrast to predictions
of previous theories. I-SALT can also deal with injection into multimode lasers and predicts partially locked
as well as the usual fully locked and free-running states.  The theory is designed to treat complex cavity geometries
and can easily incorporate different spatial injection profiles into higher dimensional cavities.

The accuracy and stability of the I-SALT theory was confirmed by excellent agreement with brute force FDTD 
simulations for the case of one-dimensional cavities. However I-SALT requires substantially less computational
effort. For example, I-SALT takes close to an hour of computational time on modern CPUs to generate the curves seen in Fig.\ \ref{p_l} while FDTD requires 168 days. In general, exact FDTD studies
of multimode lasing are computationally very demanding and could not be performed in more realistic
structures, whereas SALT has been shown to be computationally tractable in complex two-dimensional
structures such as photonic crystals \cite{chua11} and random lasers \cite{tureci08}, and 3D vectorial codes have been developed \cite{esterhazy14}.  Since I-SALT is essentially of the same degree of computational
complexity as SALT, it can be used to predict the effect of injecting signals into such complex modern laser system.
This opens up new possibilities for the study of frequency control of both single and multimode lasing, and for more
general investigations of injected systems. Finally, through comparison between the Adler theory and the
approximations made upon I-SALT to rederive this theory, it should be possible to use I-SALT to
provide an excellent ansatz to recover time-domain equations similar to those used by 
previous authors \cite{oppo_pushing_86, solari_94, zimmermann_global_01} for cavities with a complex inner structure, though some
care will need to be taken as the spatial profile of the injected signal is a very relevant parameter, as
can be seen from the simulations in section \ref{sec5b}.

\begin{acknowledgments}
We thank Stephen O'Brien, Kerry Vahala, Arthur Goestchy, and Hui Cao for helpful
discussions. This work was supported by NSF grant No. DMR-0908437. 
This work was supported in part by the facilities and staff of the Yale 
University Faculty of Arts and Sciences High Performance Computing Center.  

\end{acknowledgments}

\bibliography{references}

\end{document}